\documentclass[
    ,final            
  ]
  {aipproc}

\layoutstyle{8x11single}

\begin{document}

\title{Electromagnetic production of hyperon resonances}

\classification{13.30.Ce, 13.60.-r 14.20.Gk, 14.20.Jn}
\keywords{photoproduction, kaon, hyperon, resonance}

\author{K.~Hicks}{
  address={Dept. of Physics \& Astronomy, Ohio University, Athens OH 45701},
}

\author{D.~Keller}{
  address={Dept. of Physics \& Astronomy, Ohio University, Athens OH 45701},
}

\author{W.~Tang}{
  address={Dept. of Physics \& Astronomy, Ohio University, Athens OH 45701},
}

\begin{abstract}
The study of hyperon resonances has entered a new era of precision 
with advent of high-statistics photoproduction data from the CLAS 
detector at Jefferson Lab.  These data have multi-particle final 
states, allowing clean identification of exclusive reactions 
associated with strange mesons and baryons.  Examples of physics 
results are:  evidence for isospin interference in the decay of 
the $\Lambda(1405)$ resonance; a strong suggestion of meson cloud 
effects in the structure of the $\Sigma(1385)$ resonance; data 
from $K^*$ photoproduction that will test the existence of 
the purported $K_0(800)$ meson.  Properties of other hyperon 
resonances will also be studied in the near future.
\vspace{1pc}
\end{abstract}

\date{\today}

\maketitle

\section{Introduction}

The mass of the strange quark has a peculiar value, being close 
to the scale parameter of QCD, $\Lambda_{QCD}$ \cite{jaffe}.  The 
light quarks ($u$ and $d$) have a mass small enough so that theoretical 
tools, such as chiral perturbation theory, can be used to describe 
the dynamics of these quarks.  On the other hand, the heavy quarks 
($c$, $b$ and $t$) have a mass large enough so that non-relativistic 
potential theories, such as heavy quark effective theory (HQET), 
can be used for the dynamic of these quarks.  The strange quark 
is too heavy for chiral dynamics and too light for HQET, which 
makes it difficult to describe theoretically \cite{jaffe}.

Because of the theoretical difficulty to describe the dynamics 
of the strange quark, it is important to measure the properties 
of particles with non-zero strangeness in order to provide 
guidance for theoretical models.  For example, it has been 
debated for some time \cite{isgur} whether the structure of 
the first excited state of the $\Lambda$ baryon, the $\Lambda(1405)$ 
with $J^P = \frac{1}{2}^-$, is a $\bar{K}N$ bound state or a 
orbital excitation of the $uds$ valence quarks in a conventional 
baryon quark model.  The former would have strong 5-quark 
components in its wave-function, whereas the latter would have a
predominantly 3-quark component.  The fact that this issue is 
still not resolved is a testiment to the theoretical ambiguities 
in the dynamics of the strange quark.

Recently, the CLAS Collaboration has collected high-quality data 
for photoproduction of baryons with non-zero strangeness. The 
large statistics for these measurements forces constraints on 
theoretical models of strangeness production that were not possible 
previously.  In particular, photoproduction of the ground state 
hyperons $\Lambda$ and $\Sigma$ shows evidence for a new 
nucleon resonance ($N^*$) that couples strongly to strangeness 
decay, which may explain why it was not seen in partial wave 
analysis of pion scattering experiments.  The search for $N^*$ 
resonances is important because significantly fewer $N^*$'s 
are observed than are predicted by quark models \cite{capstick}, 
known as the ``missing" resonances problem.

In addition to data on photoproduction of the ground state hyperons, 
data on hyperon resonances (with an associated kaon) and 
on kaon resonances (with a ground state hyperon) are now being released.
These new data allow further investigations into 
the dynamics of the strange quark in mesons and baryons.

\section{CLAS Detector and Results}

The CLAS detector, located at the Thomas Jefferson National Accelerator 
Facility, is a large acceptance spectrometer with a toroidal magnetic 
field \cite{mecking}.  Beams of electrons or photons (produced via 
bremsstrahlung) with energies up to 6 GeV are incident on target of 
hydrogen, deuterium, or nuclear targets at the center of CLAS.  For 
photoproduction, the energy of the photon is given by a tagging 
detector which records the momentum of scattered electrons.  
Multi-particle final states are detected at CLAS with high rates, 
providing billions of triggered events.

Results for the reaction $\gamma p \to K^+ \Lambda$ measured at CLAS 
have been published \cite{mccracken} and are described elsewhere
\cite{rsclas}.  Of special interest, as mentioned above, 
is the observation of new $N^*$ resonances using both cross sections 
and polarization data.  Curiously, the $\Lambda$ is polarized nearly 
100\% in the direction of circularly polarized photons \cite{rsclas} 
which was not predicted by any theoretical reaction model before
the CLAS measurements.  The precision of the CLAS data has sparked 
new theoretical interest in strangeness production.

The primary focus of the current report is on hyperon resonances. 
Perhaps one of the most interesting results in the past decade from  
CLAS measurements is for the reaction $\gamma p \to K^+ \Lambda(1405)$, 
where the decays $\Lambda^* \to \Sigma^+ \pi^-$ and 
$\Lambda^* \to \Sigma^- \pi^+$ were measured simultaneously. 
In a constituent quark model, where the $\Lambda(1405)$ is described 
as an orbital excitation of the ground-state $\Lambda(1115)$, these 
two decays should be isospin symmetric.  However, in models where 
the $\Lambda(1405)$ is dynamically generated \cite{oset}, interference 
between the $\bar{K}N$ and $\Sigma\pi$ poles can produce asymmetric 
decays.  The Carnegie Melon University group at CLAS \cite{moriya} 
measured dramatically different decays 
for the $\Lambda(1405) \to \Sigma \pi$, showing that this hyperon 
resonance has a more complex structure than predicted by simple 
quark models.  This is the first time that clear isospin interference 
has been observed in the strong decay of a well established baryon 
resonance.

\subsection{$\Sigma^*(1385)$ Photoproduction }

Few data exist for photoproduction of the $\Sigma^*$ resonance, 
which is a partner to the well-known $\Delta$ resonance 
in the baryon decuplet. This is due in part to the difficulty 
of separating the $\Sigma^{*0}$ at a mass of 1.385 GeV from the 
$\Lambda(1405)$ resonance, which overlap due to their natural 
decay widths.  Hence, the first differential cross sections with 
reasonable statistics were measured for $\Sigma^{*-}$ photoproduction 
from the neutron \cite{hicks}, where the neutral $\Lambda$ 
resonances are not seen.  More precise measurements are being 
analyzed for $\Sigma^*$ photoproduction at CLAS, but are still 
very preliminary.

\begin{figure}[htb]
\includegraphics[scale=0.35]{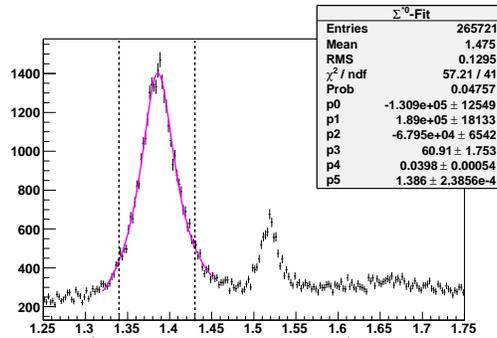}
\caption{Missing mass (in GeV) of the $K^+$ for the reaction 
$\gamma p \to K^+ \Lambda X$ showing peaks at the masses 
of the $\Sigma^0 (1385)$  and $\Lambda(1520)$ resonances.
The vertical scale shows counts.}
\label{fig:sigstar}
\end{figure}

One result of particular interest is the electromagnetic (EM) decay 
of the $\Sigma^{*0}$ to the $\Lambda \gamma$ final state, which 
was published by CLAS \cite{taylor}.  The EM decay can be calculated 
exactly for a given theoretical model, and gives direct information 
on the wave-function of the baryon.  Precise measurements have been 
made for the $\Delta \to N \gamma$ transition, yet no useful data 
were available for the $\Sigma^* \to Y \gamma$ transitions before 
the CLAS measurement.  

Using a recent high-statistics data set, 
the $\Sigma^{*0} \to \Lambda \gamma$~EM decay has been remeasured 
at CLAS, reducing the statistical uncertainty from about 30\% to 
about 10\%.  The smaller uncertainty is important, because effects 
from the pion cloud give a $\sim$30\% increase in the theoretical 
prediction \cite{bruno} for $\Delta \to N \gamma$ at low $Q^2$, 
which is necessary for agreement with data. 

The published CLAS decay width for 
$\Sigma^{*0} \to \Lambda \gamma$ is about a factor of two larger 
than theoretical predictions (for a range of theoretical models) 
\cite{taylor}, suggesting that meson cloud effects are significant.  
The new (preliminary) CLAS results 
for this EM decay width are consistent with the earlier results, 
but have much better precision.  To show the quality of the new 
data, the missing mass off the $K^+$ for the events of interest 
are shown in Fig. \ref{fig:sigstar}. Some background is seen 
under the $\Sigma^{*0}$ peak at 1.385 GeV, which is mostly 
eliminated by a cut on the confidence level of a kinematic 
fit to the final state. The key step in the analysis is to 
separate the strong decay $\Sigma^{*0} \to \Lambda \pi^0$ from 
the EM decay, which can be done reliably using kinematic fits. 

Knowledge of the quark structure of the $\Sigma^*$ resonance is 
important, as it is the lowest-lying excited state of the strange 
baryons.  Being a member of the baryon decuplet, along with the 
$\Delta$ resonance, it has spin $J=3/2$.  In the simple quark 
model, the decuplet baryons are visualized as a purely symmetric 
spin state of three valence quarks.  Relationships between the 
decuplet baryons can be calculated using isospin ($I$-spin) and 
the analogue $U$-spin $SU(3)$ flavor symmetries.  However, it is 
known that $SU(3)$ symmetries are broken, and the degree of 
symmetry breaking is something that experiments can measure. 
A broader program of measuring EM decays from a variety of the 
decuplet baryons with strangeness is in progress at CLAS, and 
will provide a window into the quark substructure of these baryons.

\begin{figure}[htb]
\includegraphics[scale=0.5]{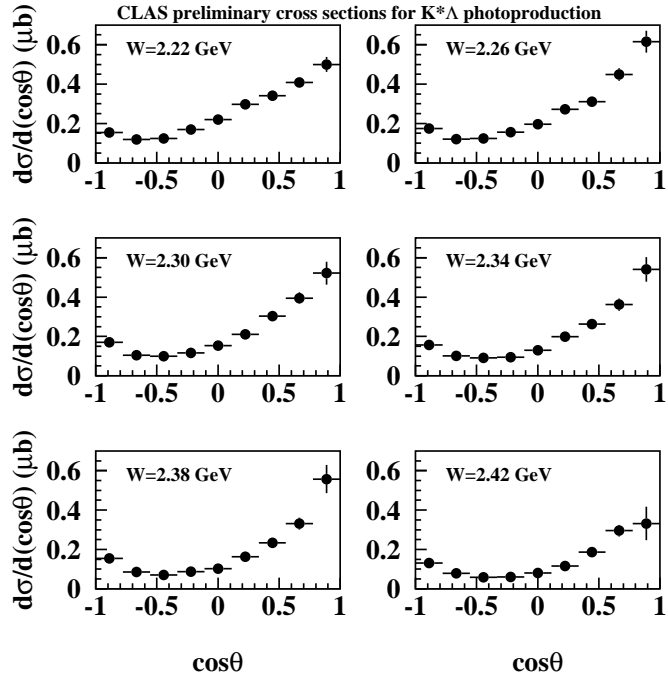}
\caption{Preliminary angular distributions for photoproduction of the 
reaction $\gamma p \to K^{*+} \Lambda$ measured at CLAS.}
\label{fig:kstar}
\end{figure}

\subsection{$K^* Y$ Photoproduction}

Before a recent CLAS publication \cite{hleiqawi}, few data existed 
for photoproduction of the $K^*(892)$ vector meson, the spin-1 
partner to the scalar kaon.  The interest in $K^*$ photoproduction 
is, in part, to understand how spin is transferred from the photon 
to the strange quark.  In the spin-0 kaon, the polarization of 
the $\bar{s}$-quark can only be inferred through measurements of 
$\Lambda$ decay in the reaction $\gamma p \to K^+ \Lambda$.  As mentioned 
above, there appears to be nearly complete transfer of spin from 
a circularly polarized photon to the $\Lambda$.  When the kaon is 
replaced by a $K^*$, the $K^*$ now carries some of the spin, 
providing new reaction dynamics.

Preliminary differential cross sections for the 
$\gamma p \to K^{*+} \Lambda$ reaction measured at CLAS are 
shown in Fig. \ref{fig:kstar}. These angular distributions are 
for increasing photon energies near threshold, and have small 
statistical uncertainties (error bars are smaller than the symbols 
for most bins).  These are the world's first high-statistics 
data for this reaction, and theorists are now developing models 
\cite{oh} for comparision with these data.

In addition to the $K^{*+}$ cross sections, preliminary results for the 
$\Lambda$ polarization from unpolarized beam show that there is 
less induced polarization compared with $K^+ \Lambda$ photoproduction 
\cite{mccracken}. Results for $K^{*+} \Lambda$ production from 
polarized photons will be measured in the near future. 

Because the $K^*$ is a vector meson, it is possible to construct 
the spin-density matrix elements for $K^{*+}\Lambda$ photoproduction. 
Preliminary results show substantial spin effects near threshold, 
similar to those seen for $\phi$-meson photoproduction.  
This suggests that significant spin transfer effects are seen to the 
$K^*$, as predicted by theory \cite{oh}. 

Within the context of a theoretical model, diagrams corresponding to 
$t$-channel exchange of virtual mesons show that the scalar meson, 
$K_0(800)$ with $J^P = 0^+$, can contribute to the forward-angle 
cross sections and spin observables.  The existence of the $K_0(800)$ 
(also called the kappa-meson) has been under discussion for many 
years.  If it exists, it is probably the strange partner to the 
well established $a_0(980)$ and $f_0(980)$ mesons in a scalar nonet. 
The $K_0(800)$ is not seen directly as a peak in the $K\pi$ invariant 
mass, presumably due to its wide width (since it can easily ``fall 
apart" into $K\pi$ because of its $J^P=0^+$ spin-parity).
A similar argument is made for the purported $\sigma$-meson, which 
would be the singlet member of this scalar nonet.  

The $K_0(800)$ does not contribute to kaon photoproduction $t$-channel, 
so only data for $K^*$ photoproduction can support or weaken the 
case for the existence of the $K_0(800)$.  So far, the case for the 
$K_0(800)$ is supported by the decay distributions of heavy mesons 
(such as $J/\psi \to \bar{K}^{*0} K^+ \pi^-$ and 
$D^+ \to K^- \pi^+ \pi^+$, see \cite{komada}) 
but the evidence remains ambiguous.  
Data for linearly polarized beam for $K^{*0}$ photoproduction, 
which has been measured at LEPS, will provide 
an unambiguous signal of the $K_0(800)$, if it exists.  Similarly, 
the ratio of $K^{*+}$ and $K^{*0}$ photoproduction total cross 
sections as a function of photon energy is predicted to be sensitive 
to the existence of the $K_0(800)$ meson.  Analysis of these data are 
in progress.  

\section{Summary and Conclusions}

The multi-particle data from high-statistics photon beams at 
CLAS have provided a rich source of results on hyperon 
resonance production.  One example is the suggestion of a 
new nucleon resonance that couples strongly to the $K^+\Lambda$ 
final state, which was further identified by the study of 
spin-transfer to the $\Lambda$ from circularly polarized beam. 
Of particular note is the isospin interference effects seen 
in the decays of the $\Lambda(1405)$ hyperon resonance.

Models of the structure of hyperon resonances can be constrained 
by measurements of EM decays, which do not suffer from the 
ambiguities of strong decays.  Although EM decays are much 
harder to measure, due to their small branching ratios, the 
results are more interesting because of the unambiguous 
predictions within a given quark model.  Results for both 
$\Delta \to N \gamma$ and $\Sigma^{*0} \to \Lambda \gamma$ 
are larger than simple quark model predictions.  In the case 
of the $\Delta$, the pion cloud makes a significant enhancement 
to the EM transition amplitude \cite{bruno}.  Similar meson cloud 
effects are now being investigated for the $\Sigma^*$ EM decay.

For $K^*$ photoproduction, high-precision data are just now 
becoming available.  Current theoretical models are 
phenomenological, and data are necessary to constrain the 
various coupling constants in the model.  One unconstrained 
parameter is the contribution of the $K_0(800)$ scalar 
meson, which is not firmly established.  The existence of 
the $K_0(800)$ would fill in the missing places in the meson 
nonet that includes the well established $a_0(980)$ and 
$f_0(980)$ mesons, and hence it is important to measure 
$K^*$ photoproduction reactions to see whether the data 
are described with diagrams that require $K_0(800)$ exchange 
in the $t$-channel.

\begin{theacknowledgments}
We gratefully acknowledge the CLAS Collaboration and the CEBAF 
accelerator staff, without which this work would not be possible.
This work is supported in part by NSF grant PHY-0653454.
\end{theacknowledgments}

\end{document}